\documentclass[11pt]{revtex4}

\usepackage{hyperref}
\usepackage{graphicx}% Include figure files
\usepackage{dcolumn}% Align table columns on decimal point
\usepackage{color}
\usepackage{bm}% bold math
\hypersetup{dvips,dvipdfm,colorlinks=true,urlcolor=magenta,filecolor=magenta,linktoc=page,citecolor=red,linkcolor=blue,bookmarks=true}
\usepackage{amsmath}
\usepackage{amsfonts}
\usepackage{amssymb}
\usepackage{epstopdf}
\usepackage{float}

\input epsf

\begin{document}

\title{Dynamical System Analysis of a Three Fluid Cosmological Model : An Invariant Manifold approach}

\author{Subhajyoti Pal\footnote {palsubhajyoti@gmail.com}}
\affiliation{Department of Mathematics, Sister Nibedita Govt
General Degree College For Girls, Kolkata-700027, West Bengal, India.}

\author{Subenoy Chakraborty\footnote {schakraborty.math@gmail.com}}
\affiliation{Department of Mathematics, Jadavpur University, Kolkata-700032, West Bengal, India.}

%%%%%%%%%%%%%%%%%%%%%%%%%%%%%%%%%%%%%%%%%%%%%%%%%%%%%%%%%%%%%%%%%%%%%%%%%%%%%%%%%%%%%%%%%%%%%%%%%%%%%%%%%%%%%%%%%%%%%%%%%%%%

\begin{abstract}

\noindent The present paper considers a three-fluid cosmological
model consisting of noninteracting dark matter, dark energy and
baryonic matter in the background of the
Friedman-Robertson-Walker-Lema\^{\i}tre flat spacetime. It has
been assumed that the dark matter takes the form of dust whereas
the dark energy is a quintessence (real) scalar field with
exponential potential. It has been further assumed that the
baryonic matter is a perfect fluid with barotropic equation of
states. The field equations for this model takes the form of an
autonomous dynamical system after some suitable changes of
variables. Then a complete stability analysis is done considering
all possible parameter (the adiabatic index of the baryonic matter
and the parameter arising from the dark energy potential) values
and for both the cases of hyperbolic and non-hyperbolic critical
points. For non-hyperbolic critical points, the invariant manifold
theory (center manifold approach) is applied. Finally various
topologically different phase planes and vector field diagrams are
produced and the cosmological interpretation of this model is
presented.\\\\
{\bf Keywords\,:} Non-hyperbolic point, Center manifold theory, Field equations.\\\\
PACS Numbers\,: 98.80.-k, 05.45.-a, 02.40.Sf, 02.40.Tt\\\\

\end{abstract}

\maketitle

%\myclassification{04.70.Dy $-$ 04.60.Kz $-$ Black hole physics~~;\\$~~~~~~~~~~~~~$ 05.70.-a $-$ Thermodynamics~~;\\$~~~~~~~~~~~~~$  95.30.Tg $-$ Thermodynamic processes, equation of state~~;\\$~~~~~~~~~~~~~$  95.30.Sf $-$ Relativity and gravitation }

%\tableofcontents

%\newpage
%%%%%%%%%%%%%%%%%%%%%%%%%%%%%%%%%%%%%%%%%%%%%%%%%%%%%%%%%%%%%%%%%%%%%%%%%%%%%%%%%%%%%%%%%%%%%%%%%%%%%%%%%%%%%%%%%%%%%%%%%%%%
%%%%%%%%%%%%%%%%%%%%%%%%%%%%%%%%%%%%%%%%%%%%%%%%%%%%%%%%%%%%%%%%%%%%%%%%%%%%%%%%%%%%%%%%%%%%%%%%%%%%%%%%%%%%%%%%%%%%%%%%%%%%

%%%%%%%%%%%%%%%%%%%%%%%%%%%%%%%%%%%%%%%%%%%%%%%%%%%%%%%%%%%%%%%%%%%%%%%%%%
\section{Introduction} \label{intro}
%%%%%%%%%%%%%%%%%%%%%%%%%%%%%%%%%%%%%%%%%%%%%%%%%%%%%%%%%%%%%%%%%%%%%%%%%%

\noindent Recently, a large number of observational data from
various sources such as Type Ia Supernova~\cite {Rei98,Perl99},
CMB anisotropies~\cite {Sper07,Komat09}, Large Scale
Structures~\cite {Teg04,Per01} and Baryon Acoustic
Oscillations~\cite {Eisen03} suggests that we are in a spatially
flat universe which after the big bang has undergone two
accelerated expansion phases, one occurred before the radiation
dominated era and the other one started not too long ago. We are
presently in this accelerated expansion phase.

\smallskip
 This present time accelerated expansion has been
attributed to a unseen and unknown matter with very large negative
pressure called the dark energy. Analysis of different
cosmological data suggests that our universe is composed of around
$70$\% dark energy, about $25$\% dark matter and the rest accounts
for baryonic matter and radiation.

\smallskip
Very few properties of the dark energy is known~\cite {AT10}. As
far as the mathematical modelling of this exotic matter is
concerned, the simplest choice for it has been the Cosmological
Constant $\Lambda$~\cite {Wein89,Carrol01,PR03}. Although these
models are capable of explaining most of the observational data,
all of them fails to explain the coincidence problem (why
expansion is happening now and why is it accelerated) and the
fine-tuning problem (why some of the parameters take exorbitantly
high values where others are not). To resolve these issues, lots
of dynamical dark energy models has been prescribed. There the
dark energy has been modelled as a scalar field. Some scalar field
models like quintessence~\cite {RP88,CDS98}, \emph{K}-essence and
Tachyonic models have attracted lots of attention.

\smallskip
 In this paper we consider a cosmological model
consisting of non-interacting dark matter, dark energy and
baryonic matter in the background of the
Friedman-Robertson-Walker-Lema\^{\i}tre flat spacetime. The dark
matter has been assumed to take the form of dust whereas the dark
energy is assumed to be a scalar field with exponential potential.
It has been further assumed that the baryonic matter is a perfect
fluid with barotropic equation of states. All these three fluids
are assumed to be non-interacting and minimally coupled to
gravity.

\smallskip
In order to study this model qualitatively, we derive the
Einstein's field equations and the Klein-Gordan equation for the
scalar field. After some suitable changes of variables, these
equations take the form of an autonomous dynamical system. Then we
find the critical points and analyze the stability of each
critical point. We note that in ~\cite {MC15} the model like ours
have been considered. But the authors did not consider the cases
of non-hyperbolic critical points. Our results differ from them in
various ways. Firstly, we apply some of the very rich theories for
the dynamical systems, namely the Invariant Manifold and the
Center Manifold Theory ~\cite {Per91,AP90,Wig03,ASB} to compute
the center manifolds for all the non-hyperbolic critical points
and then continue to do the stability analysis for them. The other
one is that we have considered all theoretically possible values
of the parameters to do a complete analysis here, whereas in their
article, they chose only a few suitable values. Lastly, we have
presented all possible topologically different phase plane
diagrams here where they included only some of them in their
article.

\smallskip
\noindent

The motivation to do stability analysis is that after considering
all cosmological and observational constraints of data, the stable
critical points in our model may depict our present universe as a
global attractor. If they do fit with the data of present
percentages of dark energy, dark matter and baryonic matter
together with radiation in the universe then our model would
successfully describe the universe.

\smallskip
The organization of this article is as follows : The section
\ref{secI} describes the Einstein field equations, Klein-Gordan
scalar field equation and energy conservation relations for our
model. Section \ref{secII} describes the formation of an
autonomous system. The section \ref{secIII} is where we present
our work on complete stability analysis. At the end of this
section, we produce the phase plane diagrams for different
topological cases. Finally section \ref{secIV} presents the
cosmological interpretations of our results and concludes our
work.

%%%%%%%%%%%%%%%%%%%%%%%%%%%%%%%%%%%%%%%%%%%%%%%%%%%%%%%%%%%%%%%%%%%%%%%%%%
\section{Equations}\label{secI}
%%%%%%%%%%%%%%%%%%%%%%%%%%%%%%%%%%%%%%%%%%%%%%%%%%%%%%%%%%%%%%%%%%%%%%%%%%

\noindent The homogeneous and isotropic flat
Friedman-Robertson-Walker-Lema\^{\i}tre spacetime is the
background of our model. This universe is assumed to be filled up
by non-interacting dark matter, dark energy and baryon. Dark
matter is assumed to be dust with energy density $\rho_m$ and the
dark energy is assumed to be a scalar field $\phi$ with the
potential as $V(\phi)$. The density $\rho_d$ and the pressure
$p_d$ of the scalar field follows the following equations :

\begin{equation}\label{de.d} \rho_d=\frac{1}{2}\dot{\phi}^{2}+V(\phi)\end{equation}
and \begin{equation}\label{de.p}
p_d=\frac{1}{2}\dot{\phi}^{2}-V(\phi).\end{equation}

Here $\dot{}$ denotes differentiation with respect to cosmic time
$t.$

The baryonic matter is assumed to be a perfect fluid with linear
equation of state
\begin{equation}\label{bm.pd}p_b=(\nu-1)\rho_b\end{equation} where $p_b$ and
$\rho_b$ are the density and the pressure of the fluid and $\nu$
is the adiabatic index of the fluid satisfying
$\frac{2}{3}<\nu\leq 2.$ In particular $\nu=1$ and
$\nu=\frac{4}{3}$ corresponds to the dust and radiation
respectively. Here we also assume that $\nu \neq 1$ All three
matter are non-interacting and minimally coupled to gravity.

The Einstein field equations for this model is
\begin{equation}\label{efe}3H^2=k(\rho_m+\rho_d+\rho_b).\end{equation}
where $H$ is the Hubble parameter and $k=8\pi G$, where $G$ is the
gravitational constant, the speed of light has been scaled to $1.$

The Klein-Gordan equation of the scalar field is
\begin{equation}\label{kge}\ddot{\phi}+3H\dot{\phi}+\frac{dV}{d\phi}=0\end{equation}

The energy conservation relations take the following form
\begin{equation}\label{ecedm}\dot{\rho_m}+3H\rho_m=0\end{equation}
\begin{equation}\label{ecede}\dot{\rho_d}+3H(\rho_d+p_d)=0\end{equation}
\begin{equation}\label{ecebm}\dot{\rho_b}+3H(\rho_b+p_b)=0\end{equation}

From (\ref{efe}),(\ref{kge}),(\ref{ecedm}),(\ref{ecede}) and
(\ref{ecebm}) we derive
\begin{equation}\label{efed}2\dot{H}=-k(\rho_m+\rho_b+p_b+\dot{\phi}^2).\end{equation}

\smallskip
\noindent The equations (\ref{efe}),(\ref{efed}) and (\ref{kge})
are the evolution equations for this model. Next we find suitable
coordinate changes such that these evolution equations form a
system of autonomous dynamical system. This is done in the
following section.

%%%%%%%%%%%%%%%%%%%%%%%%%%%%%%%%%%%%%%%%%%%%%%%%%%%%%%%%%%%%%%%%%%%%%%%%%%
\section{The autonomous system}\label{secII}
%%%%%%%%%%%%%%%%%%%%%%%%%%%%%%%%%%%%%%%%%%%%%%%%%%%%%%%%%%%%%%%%%%%%%%%%%%

\noindent We introduce the following coordinate transformations of
variables :

\begin{equation}\label{xchange} x=\sqrt{\frac{k}{6}}\frac{\dot{\phi}}{H},\end{equation}
\begin{equation}y=\label{ychange}\sqrt{\frac{k}{3}}\frac{\sqrt{V(\phi)}}{H}\end{equation}

and the density parameters

\begin{equation}\label{romchange}\Omega_m=\frac{k\rho_m}{3H^2},\end{equation}
\begin{equation}\label{robchange}\Omega_b=\frac{k\rho_b}{3H^2},\end{equation}
\begin{equation}\label{rodchange}\Omega_d=\frac{k\rho_d}{3H^2}.\end{equation}

These coordinate changes transform the Friedmann equation
(\ref{efe}) and the equation (\ref{efed}) respectively as the
following :
\begin{equation}\label{Od}\Omega_d=x^2+y^2,\end{equation}
\begin{equation}\label{mefe} \Omega_m+\Omega_b+x^2+y^2=1\end{equation}
and
\begin{equation}\label{mefed}\dot{H}=-3H^2(x^2+\frac{\Omega_m}{2}+\nu\frac{\Omega_b}{2}).\end{equation}

As $\Omega_m$ and $\Omega_b,$ the density parameters are
non-negative real quantities, so from (\ref{mefe}) $0\leq \Omega_m
\leq 1,0\leq \Omega_b \leq 1$ and $x$ and $y$ satisfies $x^2+y^2
\leq 1.$ The strict equality is only possible if the energy
densities of the dark matter and the baryonic matter is zero.

Differentiating the equations (\ref{xchange}), (\ref{ychange}) and
(\ref{romchange}) with respect to $N$ where $N=\ln a$ ($a(t)$ is
the scale factor of the universe) and using (\ref{mefe}),
(\ref{mefed}), (\ref{kge}) and (\ref{ecedm}), (\ref{ecede}) and
(\ref{ecebm}) we derive the following autonomous dynamical system
:
\begin{eqnarray}\label{ads}\frac{dx}{dN}=3x[x^2-1+\frac{\Omega_m}{2}+\frac{\nu}{2}(1-\Omega_m-x^2-y^2)]-\sqrt{\frac{3}{2k}}\frac{1}{V}\frac{dV}{d\phi}y^2\\
\frac{dy}{dN}=y[3x^2+3\frac{\nu}{2}(1-\Omega_m-x^2-y^2)+3\frac{\Omega_m}{2}+\sqrt{\frac{3}{2k}}\frac{1}{V}\frac{dV}{d\phi}x]\\
\frac{d\Omega_m}{dN}=-3\Omega_m[1-\Omega_m-2x^2-\nu(1-\Omega_m-x^2-y^2)]\end{eqnarray}

\smallskip
\noindent We end this section by expressing the relevant
cosmological parameters in terms of the above transformed
variables as
\begin{equation}\label{Od}\Omega_d=x^2+y^2,\end{equation}
\begin{equation}\omega_d=\frac{p_d}{\rho_d}=\frac{x^2-y^2}{x^2+y^2},\end{equation}
\begin{equation}\omega_{eff}=\frac{p_d+p_b}{\rho_m+\rho_d+\rho_b}=-1+2x^2+\nu(1-\Omega_m-x^2-y^2)+\Omega_m\end{equation}
and
\begin{equation}\label{dece}q=-(1+\frac{\dot{H}}{H^2})=-[(1-3x^2-\frac{3\nu}{2})-\frac{3\Omega_m}{2}(1-\nu)+\frac{3\nu}{2}(x^2+y^2)].\end{equation}

\noindent We note that for accelerated expansion, $\omega_{eff}
\leq -\frac{1}{3}$ and $q\le 0$ are two necessary conditions.

%%%%%%%%%%%%%%%%%%%%%%%%%%%%%%%%%%%%%%%%%%%%%%%%%%%%%%%%%%%%%%%%%%%%%%%%%%
\section{Stability Analysis}\label{secIII}
%%%%%%%%%%%%%%%%%%%%%%%%%%%%%%%%%%%%%%%%%%%%%%%%%%%%%%%%%%%%%%%%%%%%%%%%%%

\noindent We start working on the stability analysis of the
dynamical system (\ref{ads}) in this section. We assume that the
potential of the scalar field representing the dark energy is
exponential, ie $\frac{1}{V}\frac{dV}{d\phi}=$ some constant. We
choose $\sqrt{\frac{3}{2k}}\frac{1}{V}\frac{dV}{d\phi}=\alpha.$
\smallskip
Then the autonomous system (\ref{ads}) transforms into :
\begin{eqnarray}\label{mads}\frac{dx}{dN}=3x[x^2-1+\frac{\Omega_m}{2}+\frac{\nu}{2}(1-\Omega_m-x^2-y^2)]-\alpha y^2\\
\frac{dy}{dN}=y[3x^2+3\frac{\nu}{2}(1-\Omega_m-x^2-y^2)+3\frac{\Omega_m}{2}+\alpha x]\\
\frac{d\Omega_m}{dN}=-3\Omega_m[1-\Omega_m-2x^2-\nu(1-\Omega_m-x^2-y^2)]\end{eqnarray}

\noindent There are ten critical points of this autonomous system.
They are listed in the table below. \noindent
\begin{table}[H]
\centering%
\caption{Critical Points}\label{tab1}
\bigskip 
\begin{tabular}{|c|c|}
  \hline
  Critical Point Name & Critical Point \\
  \hline
  $C_1$ & $(0,0,0)$ \\
  \hline
  $C_2$ & $(1,0,0)$\\
  \hline
  $C_3$ & $(0,0,1)$ \\
  \hline
  $C_4$ & $(-1,0,0)$\\
  \hline
  $C_5$ & $(-\frac{\alpha}{3},\frac{1}{3}\sqrt{9-\alpha^2},0)$ \\
  \hline
  $C_6$ & $(-\frac{\alpha}{3},-\frac{1}{3}\sqrt{9-\alpha^2},0)$ \\
  \hline
  $C_7$ & $(-\frac{3\nu}{2\alpha},\frac{3}{2\alpha}\sqrt{2\nu-\nu^2},0)$ \\
  \hline
  $C_8$ & $(-\frac{3\nu}{2\alpha},-\frac{3}{2\alpha}\sqrt{2\nu-\nu^2},0)$ \\
  \hline
  $C_9$ & $(-\frac{3}{2\alpha},\frac{3}{2\alpha},(1-\frac{9}{2\alpha^2}))$\\
  \hline
  $C_{10}$ &$(-\frac{3}{2\alpha},-\frac{3}{2\alpha},(1-\frac{9}{2\alpha^2}))$ \\
  \hline
\end{tabular}
\end{table}

\bigskip

\noindent The value of the relevant cosmological parameters for
each of the critical points are given in the following table :

\begin{table}[H]
\centering%
\caption{Values of the Different Cosmological Parameters at the Critical
Points}\label{tablast}
\bigskip
\begin{tabular}{|c|c|c|c|c|c|c|}
  \hline
  Critical Points & $\omega_d$ & $\omega_{eff}$ & $\Omega_m$ & $\Omega_d$ & $\Omega_b$ & $q$ \\
  \hline
  $C_1$ & Undefined & $\nu-1$ & $0$ & $0$ & $1$ & $\frac{3\nu}{2}-1$ \\
  \hline
  $C_2$ & $1$ & $1$ & $0$ & $1$ & $0$ & $2$ \\
  \hline
  $C_3$ & Undefined & $0$ & $1$ & $0$ & $0$ & $\frac{1}{2}$ \\
  \hline
  $C_4$ & $1$ & $1$ & $0$ & $1$ & $0$ & $2$ \\
  \hline
  $C_5$ & $\frac{2\alpha^2}{9}-1$ & $\frac{2\alpha^2}{9}-1$ & $0$ & $1$ & $0$ & $\frac{\alpha^2}{3}-1$ \\
  \hline
  $C_6$ & $\frac{2\alpha^2}{9}-1$ & $\frac{2\alpha^2}{9}-1$ & $0$ & $1$ & $0$ & $\frac{\alpha^2}{3}-1$ \\
  \hline
  $C_7$ & $\nu-1$ & $\nu-1$ & $0$ & $\frac{9\nu}{2\alpha^2}$ & $1-\frac{9\nu}{2\alpha^2}$ & $\frac{3\nu}{2}-1$ \\
  \hline
  $C_8$ & $\nu-1$ & $\nu-1$ & $0$ & $\frac{9\nu}{2\alpha^2}$ & $1-\frac{9\nu}{2\alpha^2}$ & $\frac{3\nu}{2}-1$ \\
  \hline
  $C_9$ & $0$ & $0$ & $1-\frac{9}{2\alpha^2}$ & $\frac{9}{2\alpha^2}$ & $0$ & $\frac{1}{2}$ \\
  \hline
  $C_{10}$ & $0$ & $0$ & $1-\frac{9}{2\alpha^2}$ & $\frac{9}{2\alpha^2}$ & $0$ & $\frac{1}{2}$ \\
  \hline
\end{tabular}
 \end{table}

\bigskip

\noindent For every critical point and for each of their subcases
we will do two successive change of variables to bring them to a
form with which the calculation of center manifold and dynamics of
the reduced system will be easy. For each of the cases and
subcases the transformations are as follows :

\noindent If

$X=\left(%
\begin{array}{c}
  x \\
  y \\
\Omega_m \\
\end{array}%
\right),$  $\bar{X}=\left(%
\begin{array}{c}
  \bar{x} \\
  \bar{y} \\
\bar{\Omega_m} \\
\end{array}%
\right)$ and $\bar{\bar{X}}=\left(%
\begin{array}{c}
  \bar{\bar{x}} \\
  \bar{\bar{y}}\\
\bar{\bar{\Omega_m}}\\
\end{array}%
\right),$

\noindent

\noindent then $\bar{X}=X-A$ and $\bar{\bar{X}}=P^{-1}\bar{X}$
where $A$ and $P$ is some $3\times 1$ and $3\times 3$ matrices
respectively with $P$ being non-singular. The exact form of $A$
and $P$ will vary from case to case and we will mention them while
studying each of the cases and subcases.

%%% ----------------------------------------------------------------------

\subsection{Critical Point $C_1$}

For critical point $C_1$, $A=\left(%
\begin{array}{c}
  0 \\
  0 \\
  0 \\
\end{array}%
\right)$ and the Jacobian matrix of the system (\ref{mads}) at
this critical point has the characteristic polynomial
\begin{equation}\lambda^3+(6-6\nu)\lambda^2+[\frac{3\nu(3\nu-3)}{2}+\frac{(3\nu-6)(9\nu-6)}{4}]\lambda-\frac{3\nu(3\nu-3)(3\nu-6)}{4}=0.\end{equation}
which has eigen values as $3\nu-3,\frac{3\nu}{2}-3$ and
$\frac{3\nu}{2}.$ The stability analysis when $\nu$ is not $0,1$
or $2$ (hyperbolic cases) is an easy application of the stability
analysis of the linear system and the Hartman-Gr\"{o}bman Theorem.

\smallskip

\noindent So now we will present a table containing the
non-hyperbolic subcases and the result of their stability
analysis. Later We will write the stability results for all
possible subcases for $C_1$.

\noindent Before we write down the table, we introduce a notation.
$'$ represents derivative with respect to $N.$

\bigskip
\begin{table}[H]
\centering%
\caption{$C_1$(Reduced System)}\label{tab2}
\bigskip
\begin{tabular}{|c|c|c|}
  \hline
  $\nu$ & Center Manifold & Reduced System \\
  \hline
   $0$ & {$\bar{\bar{x}}=O(\bar{\bar{\Omega_m}}^3),\bar{\bar{y}}=-\frac{\alpha}{3}\bar{\bar{\Omega_m}}^2+O(\bar{\bar{\Omega_m}}^3)$} & $\bar{\bar{\Omega_m}}'=-\frac{\alpha^2}{3}\bar{\bar{\Omega_m}}^3+O(\bar{\bar{\Omega_m}}^4)$\\
  \hline
   $1$ & $\bar{\bar{y}}=O(\bar{\bar{x}}^3),\bar{\bar{\Omega_m}}=O(\bar{\bar{x}}^3)$ & $\bar{\bar{x}}'=0$ \\
  \hline
   $2$ & $\bar{\bar{x}}=O(\bar{\bar{y}}^3),\bar{\bar{\Omega_m}}=O(\bar{\bar{y}}^3)$ & $\bar{\bar{y}}'=0$  \\
  \hline
\end{tabular}
\end{table}

For all these subcases,
$P=\left(%
\begin{array}{ccc}
  0 & 1 & 0 \\
  0 & 0 & 1 \\
  1 & 0 & 0 \\
\end{array}%
\right).$

\smallskip

We summarize our results for $C_1$ in the table below to end this
subsection.

\bigskip
\begin{table}[H]
\centering%
\caption{Summary for the critical point $C_1$}\label{tab3}
\bigskip
\begin{tabular}{|c|c|c|c|}
  \hline
  Case & $\nu$ & Stability(RS) & Stability(DS) \\
  \hline
  a & $\nu < 0$ & NA & Stable \\
  \hline
  b & $\nu=0$ & Stable & Stable \\
  \hline
  c & $0<\nu<1$ & NA & Saddle \\
  \hline
  d & $\nu=1$ & Center & Center-Saddle \\
  \hline
  e & $1<\nu<2$ & NA & Saddle \\
  \hline
  f & $\nu=2$ & Center & Center-Unstable \\
  \hline
  g & $2<\nu$ & NA & Unstable \\
  \hline
\end{tabular}
\end{table}

\noindent Here RS stands for the Reduced System, DS stands for the
whole system and NA stands for Not Applicable (Hyperbolic Cases).

\subsection{Critical Point $C_2$}

For critical point $C_2$, $A=\left(%
\begin{array}{c}
  1 \\
  0 \\
  0 \\
\end{array}%
\right)$ and the Jacobian matrix of the system (\ref{mads}) at
this critical point has the characteristic polynomial
\begin{equation}\lambda^3+(3\nu-\alpha-12)\lambda^2+[18-(3\nu-9)(\alpha+3)-9\nu]\lambda-(9\nu-18)(\alpha+3)=0.\end{equation}
which has eigen values as $3,6-3\nu$ and $\alpha+3.$ The stability
analysis when $\nu$ is not $2$ and $\alpha$ is not $-3$
(hyperbolic cases) is again an easy application of the stability
analysis of the linear system and the Hartman-Gr\"{o}bman Theorem.

\smallskip

\noindent So we will present a table containing all the
non-hyperbolic subcases and the result of their stability
analysis. Later We will write the stability results for all
possible subcases for $C_2$.

\bigskip

\begin{table}[H]
\centering%
\caption{$C_2$(Reduced System)}\label{tab4}
\bigskip
\begin{tabular}{|c|c|c|c|}
  \hline
  $\nu$ & $\alpha$ & Center Manifold & Reduced System \\
  \hline
  $2$ & $-3$ & $\bar{\bar{\Omega_m}}=O(\parallel(\bar{\bar{x}},\bar{\bar{y}})\parallel^3)$ & $\bar{\bar{x}}'=-3\bar{\bar{x}}\bar{\bar{y}}^2,\bar{\bar{y}}'=-3\bar{\bar{x}}\bar{\bar{y}}$ \\
  \hline
  $2$ & $\neq -3$ & $\bar{\bar{y}}=O(\bar{\bar{x}}^3),\bar{\bar{\Omega_m}}=O(\bar{\bar{x}}^3)$ & $\bar{\bar{x}}'=0$ \\
  \hline
  $\neq 2$ & $-3$ & $\bar{\bar{y}}=O(\bar{\bar{x}}^3),\bar{\bar{\Omega_m}}=-\frac{1}{2}\bar{\bar{x}}^2+O(\bar{\bar{x}}^3)$ & $\bar{\bar{x}}'=-\frac{3}{2}\bar{\bar{x}}^3$ \\
  \hline
\end{tabular}
\end{table}

When $\nu=2$ and $\alpha=-3$ the $P$ is $\left(%
\begin{array}{ccc}
  1 & 0 & -\frac{1}{2} \\
  0 & 1 & 0 \\
  0 & 0 & 1 \\
\end{array}%
\right).$

For $\nu=2$ and $\alpha \neq -3,$ $P=\left(%
\begin{array}{ccc}
  1 & -\frac{1}{2} & 0 \\
  0 & 0 & 1 \\
  0 & 1 & 0 \\
\end{array}%
\right).$

When $\nu \neq 2$ and $\alpha=-3$ then $P= \left(%
\begin{array}{ccc}
  0 & -\frac{1}{2} & 1 \\
  1 & 0 & 0 \\
  0 & 1 & 0 \\
\end{array}%
\right).$

\smallskip

We end this subsection by summarizing our results for $C_2$ in a
table.

\bigskip

\begin{table}[H]
\centering%
\caption{A summary for the critical point $C_2$}\label{tab5}
\bigskip
\begin{tabular}{|c|c|c|c|c|}
  \hline
  Case & $\nu$ & $\alpha$ & Stability(RS) & Stability(DS) \\
  \hline
  a & $<2$ & $<-3$ & NA &  Saddle \\
  \hline
  b & $<2$ & $-3$ & Stable & Saddle \\
  \hline
  c & $<2$ & $>-3$ &  NA & Unstable \\
  \hline
  d & $2$ & $>-3$ &  Center & Center-Unstable \\
  \hline
  e & $2$ & $-3$ & Stable & Saddle \\
  \hline
  f & $2$ & $<-3$ & Center  & Center-Saddle \\
  \hline
  g & $>2$ & $<-3$ & NA & Saddle \\
  \hline
  h & $>2$ & $-3$ & Stable & Saddle \\
  \hline
  i & $>2$ & $>-3$ & NA & Saddle \\
  \hline
\end{tabular}
\end{table}

\bigskip

\subsection{Critical Point $C_3$}

For $C_3$ the $A=\left(%
\begin{array}{c}
  0 \\
  0 \\
  1 \\
\end{array}%
\right).$

\noindent The Jacobian Matrix at $C_3$ has the characteristic
polynomial

\begin{equation}\lambda^3+(3\nu-3)\lambda^2-\frac{9}{4}\lambda+[\frac{27}{4}-\frac{27\nu}{4}]=0\end{equation}

\smallskip

\noindent which has the eigenvalues $-\frac{3}{2},\frac{3}{2}$ and
$3-3\nu.$

\noindent Hence the stability analysis is almost trivial if
$\nu\neq 1.$ Because then it is an immediate application of the
stability analysis of the linear cases and the Hartman-Gr\"{o}bman
theorem.

\noindent Therefore we assume that $\nu=1$ and proceed with our
stability analysis.

\bigskip
\begin{table}[H]
\centering%
\caption{$C_3$(Reduced System)}\label{tab6}
\bigskip
\begin{tabular}{|c|c|c|}
  \hline
  $\nu$ & Center Manifold & Reduced System \\
  \hline
  $1$ & $\bar{\bar{y}}=O(\bar{\bar{x}}^3),\bar{\bar{\Omega_m}}=O(\bar{\bar{x}}^3)$ & $\bar{\bar{x}}'=0$ \\
  \hline
\end{tabular}
\end{table}

\noindent For this subcase $P$ is $\left(%
\begin{array}{ccc}
  0 & 0 & 1 \\
  0 & 1 & 0 \\
  1 & 0 & 0 \\
\end{array}%
\right).$

\bigskip

\noindent At the end, here is the summary of the results for $C_3.$

\bigskip
\begin{table}[H]
\centering%
\caption{The critical point $C_3,$ a summary}\label{tab7}
\bigskip
\begin{tabular}{|c|c|c|c|}
  \hline
  Case & $\nu$ & Stability(RS) & Stability(DS) \\
  \hline
  a & $<1$ & NA & Saddle \\
  \hline
  b & $1$ & Center & Center-Saddle \\
  \hline
  c & $>1$ & NA & Saddle \\
  \hline
\end{tabular}
\end{table}
\bigskip

\subsection{Critical Point $C_4$}

For critical point $C_4$, $A=\left(%
\begin{array}{c}
  -1 \\
  0 \\
  0 \\
\end{array}%
\right)$ and the Jacobian matrix of the system (\ref{mads}) at
this critical point has the characteristic polynomial
\begin{equation}\lambda^3+(3\nu+\alpha-12)\lambda^2+[(3\nu-6)(\alpha-6)-3\alpha+9]\lambda-(3\nu-6)(3\alpha-9)=0.\end{equation}
which has eigen values as $3,6-3\nu$ and $3-\alpha.$ The stability
analysis when $\nu$ is not $2$ and $\alpha$ is not $3$ (hyperbolic
cases) is easy again as said before.

\smallskip

\noindent Now we will present a table containing all the
non-hyperbolic subcases and the result of their stability
analysis. Later We will write the stability results for all
possible subcases for $C_4$.

\bigskip
\begin{table}[H]
\centering%
\caption{$C_4$(reduced System)}\label{tab8}
\bigskip
\begin{tabular}{|c|c|c|c|}
  \hline
  $\nu$ & $\alpha$ & Center Manifold & Reduced System \\
  \hline
  $2$ & $3$ & $\bar{\bar{\Omega_m}}=O(\parallel(\bar{\bar{x}},\bar{\bar{y}})\parallel^3)$ & $\bar{\bar{x}}'=-3\bar{\bar{x}}\bar{\bar{y}}^2,\bar{\bar{y}}'=3\bar{\bar{x}}\bar{\bar{y}}$ \\
  \hline
  $2$ & $\neq 3$ & $\bar{\bar{y}}=O(\bar{\bar{x}}^3),\bar{\bar{\Omega_m}}=O(\bar{\bar{x}}^3)$ & $\bar{\bar{x}}'=0$ \\
  \hline
  $\neq 2$ & $3$ & $\bar{\bar{y}}=O(\bar{\bar{x}}^3),\bar{\bar{\Omega_m}}=\frac{1}{2}\bar{\bar{x}}^2+O(\bar{\bar{x}}^3)$ & $\bar{\bar{x}}'=-\frac{3}{2}\bar{\bar{x}}^3$ \\
  \hline
\end{tabular}
\end{table}

\smallskip

When $\nu=2$ and $\alpha=3$ the $P$ is $\left(%
\begin{array}{ccc}
  1 & 0 & \frac{1}{2} \\
  0 & 1 & 0 \\
  0 & 0 & 1 \\
\end{array}%
\right).$

For $\nu=2$ and $\alpha \neq 3,$ $P=\left(%
\begin{array}{ccc}
  1 & \frac{1}{2} & 0 \\
  0 & 0 & 1 \\
  0 & 1 & 0 \\
\end{array}%
\right).$

When $\nu \neq 2$ and $\alpha=3$ then $P= \left(%
\begin{array}{ccc}
  0 & \frac{1}{2} & 1 \\
  1 & 0 & 0 \\
  0 & 1 & 0 \\
\end{array}%
\right).$

\smallskip

We end this subsection by summarizing our results for $C_4$ in a
table.

\smallskip
\begin{table}[H]
\centering%
\caption{$C_4,$ a stability analysis}\label{tab9}
\bigskip
\begin{tabular}{|c|c|c|c|c|}
  \hline
  Case & $\nu$ & $\alpha$ & Stability(RS) & Stability(DS) \\
  \hline
  a & $<2$ & $<3$ & NA &  Unstable \\
  \hline
  b & $<2$ & $3$ & Stable & Saddle \\
  \hline
  c & $<2$ & $>3$ &  NA & Saddle \\
  \hline
  d & $2$ & $>3$ &  Center & Center-Saddle \\
  \hline
  e & $2$ & $3$ & Stable & Saddle \\
  \hline
  f & $2$ & $<3$ & Center  & Center-Unstable \\
  \hline
  g & $>2$ & $<3$ & NA & Saddle \\
  \hline
  h & $>2$ & $3$ & Stable & Saddle \\
  \hline
  i & $>2$ & $>3$ & NA & Saddle \\
  \hline
\end{tabular}
\end{table}

\bigskip

\subsection{Critical Point $C_5$}

Here it is necessary that $\mid \alpha \mid\leq 3.$

\noindent $A$=$\left(%
\begin{array}{c}
  -\frac{\alpha}{3} \\
  \frac{1}{3}\sqrt{9-\alpha^2}\\
  $0$ \\
\end{array}%
\right)$

\noindent for this case.

\noindent The Jacobian matrix of the system (\ref{mads}) at this
critical point has the characteristic polynomial

\smallskip

\begin{equation}\lambda^3+(3\nu-\frac{5}{3}\alpha^2+6)\lambda^2+[18\nu-3\nu\alpha^2-7\alpha^2+\frac{8}{9}\alpha^4+9]\lambda-\frac{(9\nu-2\alpha^2)(2\alpha^4-27\alpha^2+81)}{27}=0.\end{equation}
This characteristic polynomial has eigenvalues as
$\frac{2\alpha^3}{3}-3\nu, \frac{\alpha^2}{3}-3$ and
$\frac{2\alpha^3}{3}-3.$ The stability analysis when $\nu\neq
\frac{2\alpha^2}{9}$ and $\alpha\neq \pm 3,\pm \frac{3}{\sqrt{2}}$
(hyperbolic cases) is easy as said in the above subsections.

\smallskip

\noindent Now we will present a table containing all the
non-hyperbolic subcases and the result of their stability
analysis. Since for $C_5,$ the total number of all possible
subcases is too many, we would not present the results for the
complete case in a table at the end as we did in the previous
subsections. We would rather present a diagram in $\alpha-\nu$
plane to show the stability analysis for all possible values of
$\alpha$'s and $\nu$'s.

\bigskip
\begin{table}[H]
\centering%
\caption{$C_5$(Reduced System)}\label{tab10}
\bigskip
\begin{tabular}{|c|c|c|c|}
  \hline
  $\nu$ & $\alpha$ & Center Manifold & Reduced System \\
  \hline
  $2$ & $3$ & id to subcase (e) of $C_4$ & id to subcase (e) of $C_4$ \\
  \hline
  $\neq 2$  & $3$ & id to subcase (b) and (h) of $C_4$ &  id to subcase (b) and (h) of $C_4$\\
  \hline
  $2$ & $-3$ & id to subcase (e) of $C_2$  & id to subcase (e) of $C_2$ \\
  \hline
  $\neq 2$ & $-3$ & id to subcase (b) and (h) of $C_2$ & id to subcase (b) and (h) of $C_2$ \\
  \hline
  $1$ & $\frac{3}{\sqrt{2}}$ & $\bar{\bar{\Omega_m}}=\frac{1}{\sqrt{2}}\bar{\bar{x}}^2+O(\parallel(\bar{\bar{x}},\bar{\bar{y}})\parallel^3)$ & $\bar{\bar{x}}'=-3\sqrt{2}\bar{\bar{x}}^2,\bar{\bar{y}}'=-3\sqrt{2}\bar{\bar{x}}\bar{\bar{y}}$ \\
  \hline
  $\neq 1$ & $\frac{3}{\sqrt{2}}$ & $\bar{\bar{x}}=r\bar{\bar{y}}^2+O(\bar{\bar{y}}^3),\bar{\bar{\Omega_m}}=s\bar{\bar{y}}^2+O(\bar{\bar{y}}^3)$ & $\bar{\bar{y}}'=-3\bar{\bar{y}}^2$ \\
  \hline
  $1$ & $-\frac{3}{\sqrt{2}}$ & $\bar{\bar{\Omega_m}}=\frac{1}{\sqrt{2}}\bar{\bar{x}}^2+O(\parallel(\bar{\bar{x}},\bar{\bar{y}})\parallel^3)$ & $\bar{\bar{x}}'=3\sqrt{2}\bar{\bar{x}}^2,\bar{\bar{y}}'=3\sqrt{2}\bar{\bar{x}}\bar{\bar{y}}$ \\
  \hline
  $\neq 1$ & $-\frac{3}{\sqrt{2}}$ & $\bar{\bar{x}}=r\bar{\bar{y}}^2+O(\bar{\bar{y}}^3),\bar{\bar{\Omega_m}}=s\bar{\bar{y}}^2+O(\bar{\bar{y}}^3)$ & $\bar{\bar{y}}'=-3\bar{\bar{y}}^2$ \\
  \hline
  $\frac{2\alpha^2}{9}$ & $\alpha$ & $\bar{\bar{y}}=t\bar{\bar{x}}^2+O(\bar{\bar{x}}^3),\bar{\bar{\Omega_m}}=O(\bar{\bar{x}}^3)$ & $\bar{\bar{x}}'=-u\bar{\bar{x}}^2+O(\bar{\bar{x}}^3)$ \\
  \hline
  $0$ & $0$ & $\bar{\bar{y}}=O(\bar{\bar{x}}^3),\bar{\bar{\Omega_m}}=O(\bar{\bar{x}}^3)$ & $\bar{\bar{x}}'=O(\bar{\bar{x}}^3)$ \\
  \hline
\end{tabular}
\end{table}

\noindent Where "id" means "identical" and
$r=\frac{\sqrt{2}(3\nu-4)}{4(2\nu-3)}$,
 $s=\frac{\sqrt{2}(1-\nu)}{4(2\nu-3)}$,
$t=\frac{3\sqrt{9-\alpha^2}(9\alpha^2-4\alpha^4)}{4\alpha^6-72\alpha^4+405\alpha^2-729}$
and $u=-\frac{4\alpha^2\sqrt{9-\alpha^2}}{2\alpha^2-9}.$ We also
note that in the subcase before the last one in the table, the
ordered pair $(\nu,\alpha)$ is not permitted to take the values
$(2,3),(2,-3),(1,\frac{3}{\sqrt{2}})$ and
$(1,-\frac{3}{\sqrt{2}}).$

\noindent When $\nu=1$ and $\alpha=\frac{3}{\sqrt{2}}$ the $P$ is $\left(%
\begin{array}{ccc}
  1 & 0 & 1 \\
  0 & 0 & 1 \\
  0 & 1 & 0 \\
\end{array}%
\right).$

\noindent For $\nu\neq 1$ and $\alpha=\frac{3}{\sqrt{2}},$ $P=\left(%
\begin{array}{ccc}
  1 & \frac{1}{\sqrt{2}} & \frac{2-\nu}{\nu-1} \\
  1 & 0 & 1 \\
  0 & 1 & 0 \\
\end{array}%
\right).$

\noindent For $\nu=1$ and $\alpha=-\frac{3}{\sqrt{2}}$ the $P$ is $\left(%
\begin{array}{ccc}
  1 & 0 & -1 \\
  0 & 0 & 1 \\
  0 & 1 & 0 \\
\end{array}%
\right)$

\noindent and when  $\nu\neq 1$ and $\alpha=-\frac{3}{\sqrt{2}},$ $P=\left(%
\begin{array}{ccc}
  -1 & -\frac{1}{\sqrt{2}} & -\frac{2-\nu}{\nu-1} \\
  1 & 0 & 1 \\
  0 & 1 & 0 \\
\end{array}%
\right).$

\noindent Also in the final two subcases $P=\left(%
\begin{array}{ccc}
  -\frac{2\alpha(9-\alpha^2)^{3/2}}{2\alpha^4-27\alpha^2+81} & \frac{\sqrt{9-\alpha^2}}{\alpha} & \frac{3}{2\alpha} \\
  1 & 1 & 0 \\
  0 & 0 & 1 \\
\end{array}%
\right)$

\noindent and $P=\left(%
\begin{array}{ccc}
  1 & 0 & 0 \\
  0 & -\frac{1}{2} & 1 \\
  0 & 1 & 0 \\
\end{array}%
\right)$

\noindent respectively.

\smallskip

\noindent Since we will find that the stability diagram for $C_5$
and $C_6$ is identical, it will be presented at the end of the
next subsection. We now proceed with the next subsection.

\bigskip

\subsection{Critical Point $C_6$}

Here it is also necessary that $\mid \alpha \mid\leq 3.$

\noindent $A$=$\left(%
\begin{array}{c}
  -\frac{\alpha}{3} \\
  -\frac{1}{3}\sqrt{9-\alpha^2}\\
  $0$ \\
\end{array}%
\right)$
\noindent in this case.

\noindent The Jacobian matrix of the system (\ref{mads}) at this
critical point has the characteristic polynomial

\smallskip

\begin{equation}\lambda^3+(3\nu-\frac{5}{3}\alpha^2+6)\lambda^2+[18\nu-3\nu\alpha^2-7\alpha^2+\frac{8}{9}\alpha^4+9]\lambda+\frac{(9\nu-2\alpha^2)(2\alpha^4-27\alpha^2+81)}{27}=0.\end{equation}
This characteristic polynomial has eigenvalues as
$\frac{2\alpha^3}{3}-3\nu, \frac{\alpha^2}{3}-3$ and
$\frac{2\alpha^3}{3}-3.$ The stability analysis when $\nu\neq
\frac{2\alpha^2}{9}$ and $\alpha\neq \pm 3,\pm \frac{3}{\sqrt{2}}$
(hyperbolic cases) is again easy.

\smallskip

\noindent Now we will present a table containing all the
non-hyperbolic subcases and the result of their stability
analysis. For $C_6$ too the total number of all possible subcases
is many, so we would not present the results for the complete case
in a table. We would rather present a diagram in $\alpha-\nu$
plane to show the stability analysis for all possible values of
$\alpha$'s and $\nu$'s as we did for $C_5$.

\bigskip
\begin{table}[H]
\centering%
\caption{$C_6$(Reduced System)}\label{tab11}
\bigskip
\begin{tabular}{|c|c|c|c|}
  \hline
  $\nu$ & $\alpha$ & Center Manifold & Reduced System \\
  \hline
  $2$ & $3$ & id to subcase (e) of $C_4$ & id to subcase (e) of $C_4$ \\
  \hline
  $\neq 2$  & $3$ & id to subcase (b) and (h) of $C_4$ &  id to subcase (b) and (h) of $C_4$\\
  \hline
  $2$ & $-3$ & id to subcase (e) of $C_2$  & id to subcase (e) of $C_2$ \\
  \hline
  $\neq 2$ & $-3$ & id to subcase (b) and (h) of $C_2$ & id to subcase (b) and (h) of $C_2$ \\
  \hline
  $1$ & $\frac{3}{\sqrt{2}}$ & $\bar{\bar{\Omega_m}}=-\frac{1}{\sqrt{2}}\bar{\bar{x}}^2+O(\parallel(\bar{\bar{x}},\bar{\bar{y}})\parallel^3)$ & $\bar{\bar{x}}'=-3\sqrt{2}\bar{\bar{x}}^2,\bar{\bar{y}}'=-3\sqrt{2}\bar{\bar{x}}\bar{\bar{y}}$ \\
  \hline
  $\neq 1$ & $\frac{3}{\sqrt{2}}$ & $\bar{\bar{x}}=-r\bar{\bar{y}}^2+O(\bar{\bar{y}}^3),\bar{\bar{\Omega_m}}=-s\bar{\bar{y}}^2+O(\bar{\bar{y}}^3)$ & $\bar{\bar{y}}'=-3\bar{\bar{y}}^2$ \\
  \hline
  $1$ & $-\frac{3}{\sqrt{2}}$ & $\bar{\bar{\Omega_m}}=-\frac{1}{\sqrt{2}}\bar{\bar{x}}^2+O(\parallel(\bar{\bar{x}},\bar{\bar{y}})\parallel^3)$ & $\bar{\bar{x}}'=3\sqrt{2}\bar{\bar{x}}^2,\bar{\bar{y}}'=3\sqrt{2}\bar{\bar{x}}\bar{\bar{y}}$ \\
  \hline
  $\neq 1$ & $-\frac{3}{\sqrt{2}}$ & $\bar{\bar{x}}=-r\bar{\bar{y}}^2+O(\bar{\bar{y}}^3),\bar{\bar{\Omega_m}}=-s\bar{\bar{y}}^2+O(\bar{\bar{y}}^3)$ & $\bar{\bar{y}}'=-3\bar{\bar{y}}^2$ \\
  \hline
  $\frac{2\alpha^2}{9}$ & $\alpha$ & $\bar{\bar{y}}=-t\bar{\bar{x}}^2+O(\bar{\bar{x}}^3),\bar{\bar{\Omega_m}}=O(\bar{\bar{x}}^3)$ & $\bar{\bar{x}}'=u\bar{\bar{x}}^2+O(\bar{\bar{x}}^3)$ \\
  \hline
  $0$ & $0$ & $\bar{\bar{y}}=O(\bar{\bar{x}}^3),\bar{\bar{\Omega_m}}=O(\bar{\bar{x}}^3)$ & $\bar{\bar{x}}'=O(\bar{\bar{x}}^3)$ \\
  \hline
\end{tabular}
\end{table}

\noindent Where $r,s,t$ and $u$ take usual values as defined in
the subsection of $C_5.$ As like in the previous subsection here
is also in the subcase before the last one in the table, the
ordered pair $(\nu,\alpha)$ is not permitted to take the values
$(2,3),(2,-3),(1,\frac{3}{\sqrt{2}})$ and
$(1,-\frac{3}{\sqrt{2}}).$

\noindent When $\nu=1$ and $\alpha=\frac{3}{\sqrt{2}},$ $P=\left(%
\begin{array}{ccc}
  1 & 0 & -1 \\
  0 & 0 & 1 \\
  0 & 1 & 0 \\
\end{array}%
\right).$

\noindent For $\nu\neq 1$ and $\alpha=\frac{3}{\sqrt{2}},$ $P=\left(%
\begin{array}{ccc}
  -1 & \frac{1}{\sqrt{2}} & -\frac{2-\nu}{\nu-1} \\
  1 & 0 & 1 \\
  0 & 1 & 0 \\
\end{array}%
\right).$

\noindent For $\nu=1$ and $\alpha=-\frac{3}{\sqrt{2}},$ $P=\left(%
\begin{array}{ccc}
  1 & 0 & 1 \\
  0 & 0 & 1 \\
  0 & 1 & 0 \\
\end{array}%
\right)$

\noindent and when  $\nu\neq 1$ and $\alpha=-\frac{3}{\sqrt{2}},$ $P=\left(%
\begin{array}{ccc}
  1 & -\frac{1}{\sqrt{2}} & \frac{2-\nu}{\nu-1} \\
  1 & 0 & 1 \\
  0 & 1 & 0 \\
\end{array}%
\right).$

\noindent Also in the final two subcases $P=\left(%
\begin{array}{ccc}
  \frac{2\alpha(9-\alpha^2)^{3/2}}{2\alpha^4-27\alpha^2+81} & -\frac{\sqrt{9-\alpha^2}}{\alpha} & \frac{3}{2\alpha} \\
  1 & 1 & 0 \\
  0 & 0 & 1 \\
\end{array}%
\right)$

\noindent and $P=\left(%
\begin{array}{ccc}
  1 & 0 & 0 \\
  0 & -\frac{1}{2} & 1 \\
  0 & 1 & 0 \\
\end{array}%
\right)$

\noindent respectively.

\smallskip

\noindent We finish this subsection by producing the stability
diagram for both $C_5$ and $C_6$ in $\alpha-\nu$ plane in the
following :

\begin{figure}[H]
\begin{center}
\includegraphics[scale=0.5]{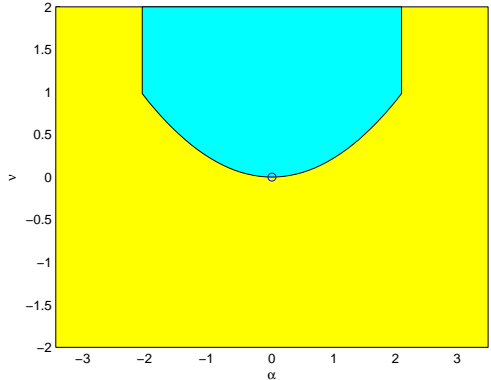}
\caption{Stability Diagram ($C_5,C_6$)}\label{sfig1}
\end{center}
\end{figure}

\noindent Here the cyan shaded area ie the area bounded by
$\alpha=\frac{3}{\sqrt{2}},\alpha=-\frac{3}{\sqrt{2}}$ and
$\nu=\frac{2\alpha^2}{9}$ represents the $\alpha-\nu$ pairs for
which (\ref{mads}) is stable. The yellow shaded region represents
parameter values for which (\ref{mads}) is saddle. Lastly the
origin is center-stable here.

\bigskip

\subsection{Critical Point $C_7$}

In this case, it is necessary that $0 \leq \nu \leq 2$ and
$\alpha\neq 0.$

\noindent $A$ takes the form $\left(%
\begin{array}{c}
  -\frac{3\nu}{2\alpha} \\
  \frac{3\sqrt{\nu(2-\nu)}}{2\alpha}\\
  $0$ \\
\end{array}%
\right).$

\noindent The Jacobian matrix of the system (\ref{mads}) at $C_7$
has the following characteristic polynomial:

\smallskip

\begin{equation}\lambda^3+(6-\frac{9\nu}{2})\lambda^2+[\frac{9(\nu-2)(9\nu^2-2\alpha^2)}{4\alpha^2}]\lambda-\frac{27\nu(9\nu-2\alpha^2)(\nu^2-3\nu+2)}{4\alpha^2}=0.\end{equation}
It has eigenvalues as $3(\nu-1),
\frac{3}{4\alpha}(\nu\alpha-2\alpha+\sqrt{(2-\nu)(36\nu^2-9\nu\alpha^2+2\alpha^2)})$
and
$\frac{3}{4\alpha}(\nu\alpha-2\alpha-\sqrt{(2-\nu)(36\nu^2-9\nu\alpha^2+2\alpha^2)}).$
The stability analysis when $\nu\neq 1, \nu\neq
\frac{2\alpha^2}{9}$ and $\nu\neq 2,\nu\neq 0$ (hyperbolic cases)
are easy by application of linear stability analysis and reduction
of non-linear case to linear case under some specific conditions.

\smallskip

\noindent So we present a table containing all the non-hyperbolic
subcases. For $C_7$ also, we will present a diagram in
$\alpha-\nu$ plane to show the stability analysis for all possible
values of $\alpha$'s and $\nu$'s at the end.

\bigskip
\begin{table}[H]
\centering%
\caption{$C_7$(Reduced System)}\label{tab12}
\bigskip
\begin{tabular}{|c|c|c|c|}
  \hline
  $\nu$ & $\alpha$ & Center Manifold & Reduced System \\
  \hline
  $1$ & $\frac{3}{\sqrt{2}}$ & id to subcase (e) of $C_5$ & id to subcase (e) of $C_5$ \\
  \hline
  $1$  & $-\frac{3}{\sqrt{2}}$ & id to subcase (g) of $C_6$ &  id to subcase (g) of $C_6$\\
  \hline
  $\neq 1$ & $\nu=\frac{2\alpha^2}{9},\alpha> 0$ & id to subcase (i) of $C_5$  & id to subcase (i) of $C_5$ \\
  \hline
  $\neq 1$ & $\nu=\frac{2\alpha^2}{9},\alpha< 0$ & id to subcase (i) of $C_6$ & id to subcase (i) of $C_6$ \\
  \hline
  $1$ & $\nu\neq \frac{2\alpha^2}{9}$ & $\bar{\bar{y}}=O(\bar{\bar{x}}^3),\bar{\bar{\Omega_m}}=O(\bar{\bar{x}}^3)$ & $\bar{\bar{x}}'=0$ \\
  \hline
  $2$ & $\alpha$ & $\bar{\bar{x}}=O(\parallel(\bar{\bar{y}},\bar{\bar{\Omega_m}})\parallel^3)$ & $\bar{\bar{y}}'=\alpha\bar{\bar{y}}\bar{\bar{\Omega_m}},\bar{\bar{\Omega_m}}'=-v\bar{\bar{y}}^2$ \\
  \hline
  $0$ & $\alpha$ & id to subcase (a) of $C_1$ & id to subcase (a) of $C_1$ \\
  \hline
\end{tabular}
\end{table}

\noindent where $v=\frac{4\alpha^2-36}{4\alpha}.$

\noindent For the last two subcases $P$ is $\left(%
\begin{array}{ccc}
  0 & w & -w \\
  0 & 1 & 1 \\
  1 & 0 & 0 \\
\end{array}%
\right)$ and $\left(%
\begin{array}{ccc}
  \frac{3}{2\alpha} & 1 & 0 \\
  0 & 0 & 1 \\
  1 & 0 & 0 \\
\end{array}%
\right)$

\noindent respectively with
$w=\frac{\alpha\sqrt{36-\alpha^2}-\alpha^2+9}{2\alpha^2-9}.$

\smallskip

\noindent Again we will find that the stability diagram for $C_7$
is exactly same as $C_8.$ Hence it will be presented at the end of
the next subsection.

\bigskip

\subsection{Critical Point $C_8$}

In this case also, it is necessary that $0 \leq \nu \leq 2$ and
$\alpha\neq 0$ and

\noindent $A$ takes the form $\left(%
\begin{array}{c}
  -\frac{3\nu}{2\alpha} \\
  -\frac{3\sqrt{\nu(2-\nu)}}{2\alpha}\\
  $0$ \\
\end{array}%
\right).$

\noindent The Jacobian matrix of the system (\ref{mads}) at $C_8$
has the characteristic polynomial:

\smallskip

\begin{equation}\lambda^3+(6-\frac{9\nu}{2})\lambda^2+[\frac{9(\nu-2)(9\nu^2-2\alpha^2)}{4\alpha^2}]\lambda-\frac{27\nu(9\nu-2\alpha^2)(\nu^2-3\nu+2)}{4\alpha^2}=0.\end{equation}
This polynomial has eigenvalues as $3(\nu-1),
\frac{3}{4\alpha}(\nu\alpha-2\alpha+\sqrt{(2-\nu)(36\nu^2-9\nu\alpha^2+2\alpha^2)})$
and
$\frac{3}{4\alpha}(\nu\alpha-2\alpha-\sqrt{(2-\nu)(36\nu^2-9\nu\alpha^2+2\alpha^2)}).$
The stability analysis when $\nu\neq 1, \nu\neq
\frac{2\alpha^2}{9}$ and $\nu\neq 2,\nu\neq 0$ (hyperbolic cases)
are easy again.
\smallskip

\noindent So we present a table containing all the non-hyperbolic
subcases for $C_8$. For this case also, at the end we will present
a diagram in $\alpha-\nu$ plane to show the stability analysis for
all possible values of $\alpha$'s and $\nu$'s.

\bigskip

\begin{table}[H]
\centering%
\caption{$C_8$(Reduced System)}\label{tab13}
\bigskip
\begin{tabular}{|c|c|c|c|}
  \hline
  $\nu$ & $\alpha$ & Center Manifold & Reduced System \\
  \hline
  $1$ & $\frac{3}{\sqrt{2}}$ & id to subcase (e) of $C_6$ & id to subcase (e) of $C_6$ \\
  \hline
  $1$  & $-\frac{3}{\sqrt{2}}$ & id to subcase (g) of $C_5$ &  id to subcase (g) of $C_5$\\
  \hline
  $\neq 1$ & $\nu=\frac{2\alpha^2}{9},\alpha> 0$ & id to subcase (i) of $C_6$  & id to subcase (i) of $C_6$ \\
  \hline
  $\neq 1$ & $\nu=\frac{2\alpha^2}{9},\alpha< 0$ & id to subcase (i) of $C_5$ & id to subcase (i) of $C_5$ \\
  \hline
  $1$ & $\nu\neq \frac{2\alpha^2}{9}$ & $\bar{\bar{y}}=O(\bar{\bar{x}}^3),\bar{\bar{\Omega_m}}=O(\bar{\bar{x}}^3)$ & $\bar{\bar{x}}'=0$ \\
  \hline
  $2$ & $\alpha$ & $\bar{\bar{x}}=O(\parallel(\bar{\bar{y}},\bar{\bar{\Omega_m}})\parallel^3)$ & $\bar{\bar{y}}'=\alpha\bar{\bar{y}}\bar{\bar{\Omega_m}},\bar{\bar{\Omega_m}}'=-v\bar{\bar{y}}^2$ \\
  \hline
  $0$ & $\alpha$ & id to subcase (a) of $C_1$ & id to subcase (a) of $C_1$ \\
  \hline
\end{tabular}
\end{table}

\noindent Where "$v$" takes it's value as defined for $C_7.$

\noindent For the last two subcases $P$ is $\left(%
\begin{array}{ccc}
  0 & q & j \\
  0 & 1 & 1 \\
  1 & 0 & 0 \\
\end{array}%
\right)$ and $\left(%
\begin{array}{ccc}
  \frac{3}{2\alpha} & 1 & 0 \\
  0 & 0 & 1 \\
  1 & 0 & 0 \\
\end{array}%
\right)$

\noindent respectively with
$j=\frac{\alpha\sqrt{36-\alpha^2}+\alpha^2-9}{2\alpha^2-9}$ and
$q=\frac{-\alpha\sqrt{36-\alpha^2}+\alpha^2-9}{2\alpha^2-9}.$

\smallskip

\noindent This subsection concludes with the stability diagram in
the $\alpha-\nu$ plane for both the $C_7$ and $C_8$ :

\begin{figure}[H]
\begin{center}
\includegraphics
[scale=0.5]{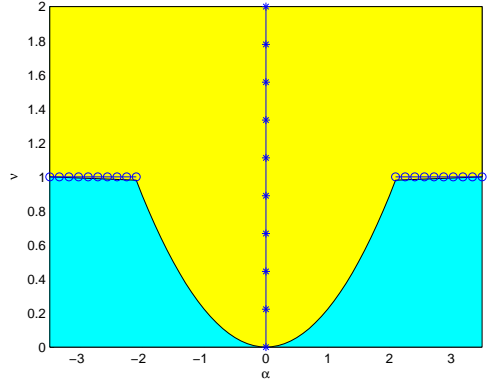}
\caption{Stability Diagram ($C_7,C_8$)}\label{sfig2}
\end{center}
\end{figure}

\noindent Here $0\leq \nu \leq 2$ and $\alpha\neq 0.$ The
cyan-shaded area ie the area bounded by $\nu=0,\nu=1$ and
$\nu=\frac{2\alpha^2}{9}$ represents stability. The yellow region
represents saddle system (\ref{mads}). The 'o' marked lines are
those parameter values for which the system (\ref{mads}) is
center-stable.

\bigskip

\subsection{Critical Point $C_9$}

In this case, it is necessary that $\alpha\neq 0.$

\noindent For this case, $A=\left(%
\begin{array}{c}
  -\frac{3\nu}{2\alpha} \\
  \frac{3\nu}{2\alpha}\\
  \frac{2\alpha^2-9}{2\alpha^2} \\
\end{array}%
\right).$

\noindent The Jacobian matrix of the system (\ref{mads}) at $C_9$
has the characteristic polynomial as following :

\smallskip

\begin{equation}\lambda^3+(3\nu-\frac{3}{2})\lambda^2+[\frac{18\nu\alpha^2-81}{4\alpha^2}]\lambda-\frac{27(2\alpha^2-9)(\nu-1)}{4\alpha^2}=0.\end{equation}
This polynomial has eigenvalues as $3(1-\nu),
\frac{-3\alpha+3\sqrt{36-7\alpha^2}}{4\alpha}$ and
$\frac{-3\alpha-3\sqrt{36-7\alpha^2}}{4\alpha}.$ The stability
analysis when $\nu\neq 1$ and $\alpha\neq
-\frac{3}{\sqrt{2}},\frac{3}{\sqrt{2}}$ (hyperbolic cases) are
easy.
\smallskip

\noindent Therefore we present a table containing all the
non-hyperbolic subcases for $C_9$. In the end we will present a
diagram in $\alpha-\nu$ plane to show the complete stability
analysis as usual.

\bigskip

\begin{table}[H]
\centering%
\caption{$C_9$(Reduced System)}\label{tab14}
\bigskip
\begin{tabular}{|c|c|c|c|}
  \hline
  $\nu$ & $\alpha$ & Center Manifold & Reduced System \\
  \hline
  $1$ & $\frac{3}{\sqrt{2}}$ & id to subcase (a) of $C_7$ & id to subcase (a) of $C_7$ \\
  \hline
  $1$  & $-\frac{3}{\sqrt{2}}$ & id to subcase (b) of $C_7$ &  id to subcase (b) of $C_7$\\
  \hline
  $\neq 1$ & $\frac{3}{\sqrt{2}}$ & id to subcase (f) of $C_5$  & id to subcase (f) of $C_5$ \\
  \hline
  $\neq 1$ & $-\frac{3}{\sqrt{2}}$ & id to subcase (h) of $C_6$ & id to subcase (h) of $C_6$ \\
  \hline
  $1$ & $\alpha\neq -\frac{3}{\sqrt{2}},\frac{3}{\sqrt{2}}$ & $\bar{\bar{y}}=O(\bar{\bar{x}}^3),\bar{\bar{\Omega_m}}=O(\bar{\bar{x}}^3)$ & $\bar{\bar{x}}'=0$ \\
  \hline
\end{tabular}
\end{table}

\smallskip

\noindent For the last subcase, $P=\left(%
\begin{array}{ccc}
  0 & \frac{\alpha}{6}+\frac{\alpha^2(\alpha-\sqrt{36-7\alpha^2})}{12(2\alpha^2-9)}
   & \frac{\alpha}{6}+\frac{\alpha^2(\alpha+\sqrt{36-7\alpha^2})}{12(2\alpha^2-9)} \\
  0 & -\frac{\alpha}{6}+\frac{\alpha^2(\alpha-\sqrt{36-7\alpha^2})}{12(2\alpha^2-9)} & -\frac{\alpha}{6}+\frac{\alpha^2(\alpha+\sqrt{36-7\alpha^2})}{12(2\alpha^2-9)} \\
  1 & 1 & 1 \\
\end{array}%
\right).$

\smallskip

\noindent \noindent We find again that the stability diagram for
$C_9$ and $C_{10}$ is totally same. Hence it will be presented at
the end of the next subsection of $C_{10}$.

\bigskip

\subsection{Critical Point $C_{10}$}

This is the last subcase. In this case too it is necessary that
$\alpha\neq 0.$

\noindent Also, $A=\left(%
\begin{array}{c}
  -\frac{3\nu}{2\alpha} \\
  \frac{3\nu}{2\alpha}\\
  \frac{2\alpha^2-9}{2\alpha^2} \\
\end{array}%
\right).$

\noindent The Jacobian matrix of the system (\ref{mads}) at
$C_{10}$ has the characteristic polynomial

\smallskip

\begin{equation}\lambda^3+(3\nu-\frac{3}{2})\lambda^2+[\frac{18\nu\alpha^2-81}{4\alpha^2}]\lambda+\frac{27(2\alpha^2-9)(\nu-1)}{4\alpha^2}=0.\end{equation}
which has eigenvalues as $3(1-\nu),
\frac{-3\alpha+3\sqrt{36-7\alpha^2}}{4\alpha}$ and
$\frac{-3\alpha-3\sqrt{36-7\alpha^2}}{4\alpha}.$ The stability
analysis when $\nu\neq 1$ and $\alpha\neq
-\frac{3}{\sqrt{2}},\frac{3}{\sqrt{2}}$ (hyperbolic cases) are
easy again.
\smallskip

\noindent Therefore we will present a table containing all the
non-hyperbolic subcases for this critical point. At the end we
will present the stability diagram in $\alpha-\nu$ plane.
\bigskip
\begin{table}[H]
\centering%
\caption{$C_{10}$(Reduced System)}\label{tab15}
\bigskip
\begin{tabular}{|c|c|c|c|}
  \hline
  $\nu$ & $\alpha$ & Center Manifold & Reduced System \\
  \hline
  $1$ & $\frac{3}{\sqrt{2}}$ & id to subcase (a) of $C_8$ & id to subcase (a) of $C_8$ \\
  \hline
  $1$  & $-\frac{3}{\sqrt{2}}$ & id to subcase (b) of $C_8$ &  id to subcase (b) of $C_8$\\
  \hline
  $\neq 1$ & $\frac{3}{\sqrt{2}}$ & id to subcase (f) of $C_6$  & id to subcase (f) of $C_6$ \\
  \hline
  $\neq 1$ & $-\frac{3}{\sqrt{2}}$ & id to subcase (h) of $C_5$ & id to subcase (h) of $C_5$ \\
  \hline
  $1$ & $\alpha\neq -\frac{3}{\sqrt{2}},\frac{3}{\sqrt{2}}$ & $\bar{\bar{y}}=O(\bar{\bar{x}}^3),\bar{\bar{\Omega_m}}=O(\bar{\bar{x}}^3)$ & $\bar{\bar{x}}'=0$ \\
  \hline
\end{tabular}
\end{table}
\smallskip

\noindent For the last subcase, $P=\left(%
\begin{array}{ccc}
  0 & \frac{\alpha}{6}+\frac{\alpha^2(\alpha-\sqrt{36-7\alpha^2})}{12(2\alpha^2-9)}
   & \frac{\alpha}{6}+\frac{\alpha^2(\alpha+\sqrt{36-7\alpha^2})}{12(2\alpha^2-9)} \\
  0 & \frac{\alpha}{6}-\frac{\alpha^2(\alpha-\sqrt{36-7\alpha^2})}{12(2\alpha^2-9)} & \frac{\alpha}{6}-\frac{\alpha^2(\alpha+\sqrt{36-7\alpha^2})}{12(2\alpha^2-9)} \\
  1 & 1 & 1 \\
\end{array}%
\right).$

\smallskip

\noindent Lastly, we provide the stability diagram in the
$\alpha-\nu$ plane for $C_9$ and $C_{10}$ and end this section :

\begin{figure}[H]
\begin{center}
\includegraphics[scale=0.5]{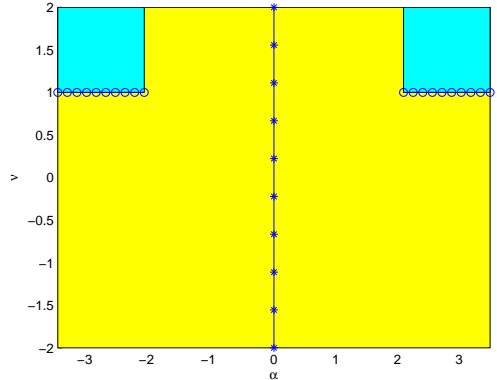}
\caption{Stability Diagram ($C_9,C_{10}$)} \label{sfig3}
\end{center}
\end{figure}

\bigskip

\noindent Here it is necessary that $\alpha\neq 0.$ The two
squares in the top corners bounded by the lines
$\nu=1,\alpha=\frac{3}{\sqrt{2}}$ and
$\nu=1,\alpha=-\frac{3}{\sqrt{2}}$ respectively are cyan-shaded.
They represent the parameter values for which the system
(\ref{mads}) is stable. The rest of the region represents saddle
system.The 'o' marked line represents center-stability.

\begin{figure}[H]
\begin{center}
\includegraphics
[scale=0.75]{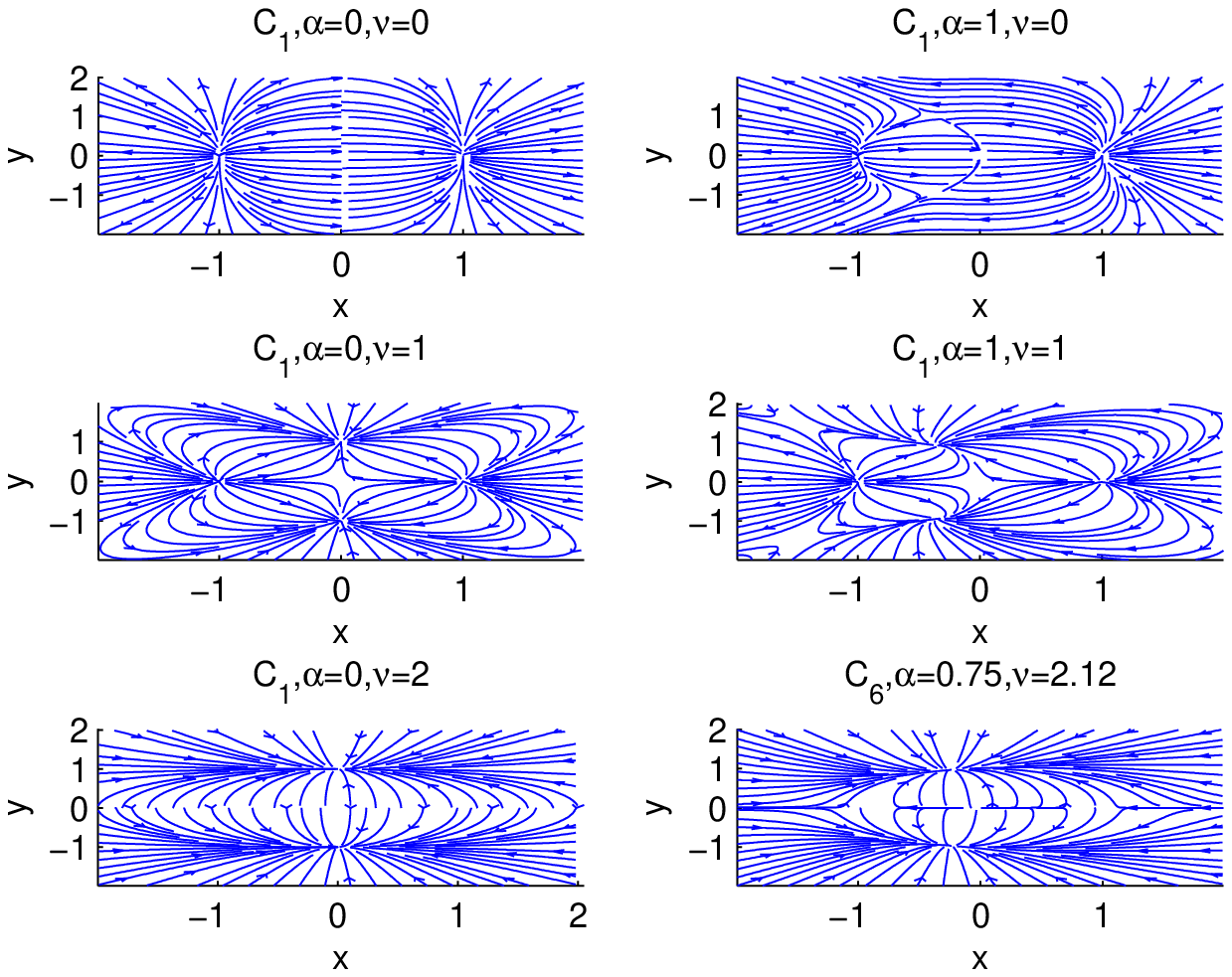}
\caption{$C_1,C_6$ }\label{fig21}
\end{center}
\end{figure}

\begin{figure}[H]
\begin{center}
\includegraphics
[scale=0.75]{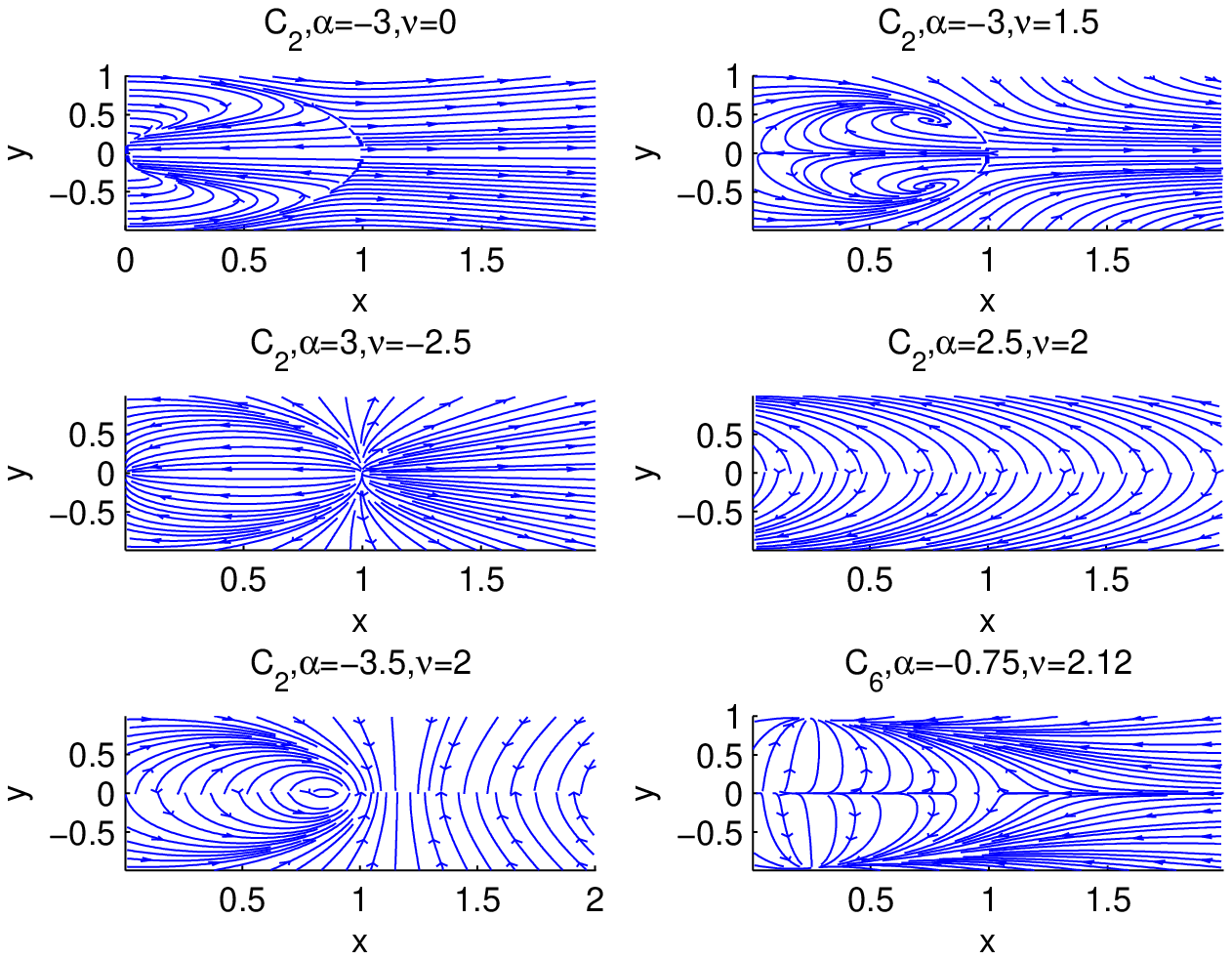}
\caption{$C_2,C_6$ }\label{fig22}
\end{center}
\end{figure}

\begin{figure}[H]
\begin{center}
\includegraphics
[scale=0.75]{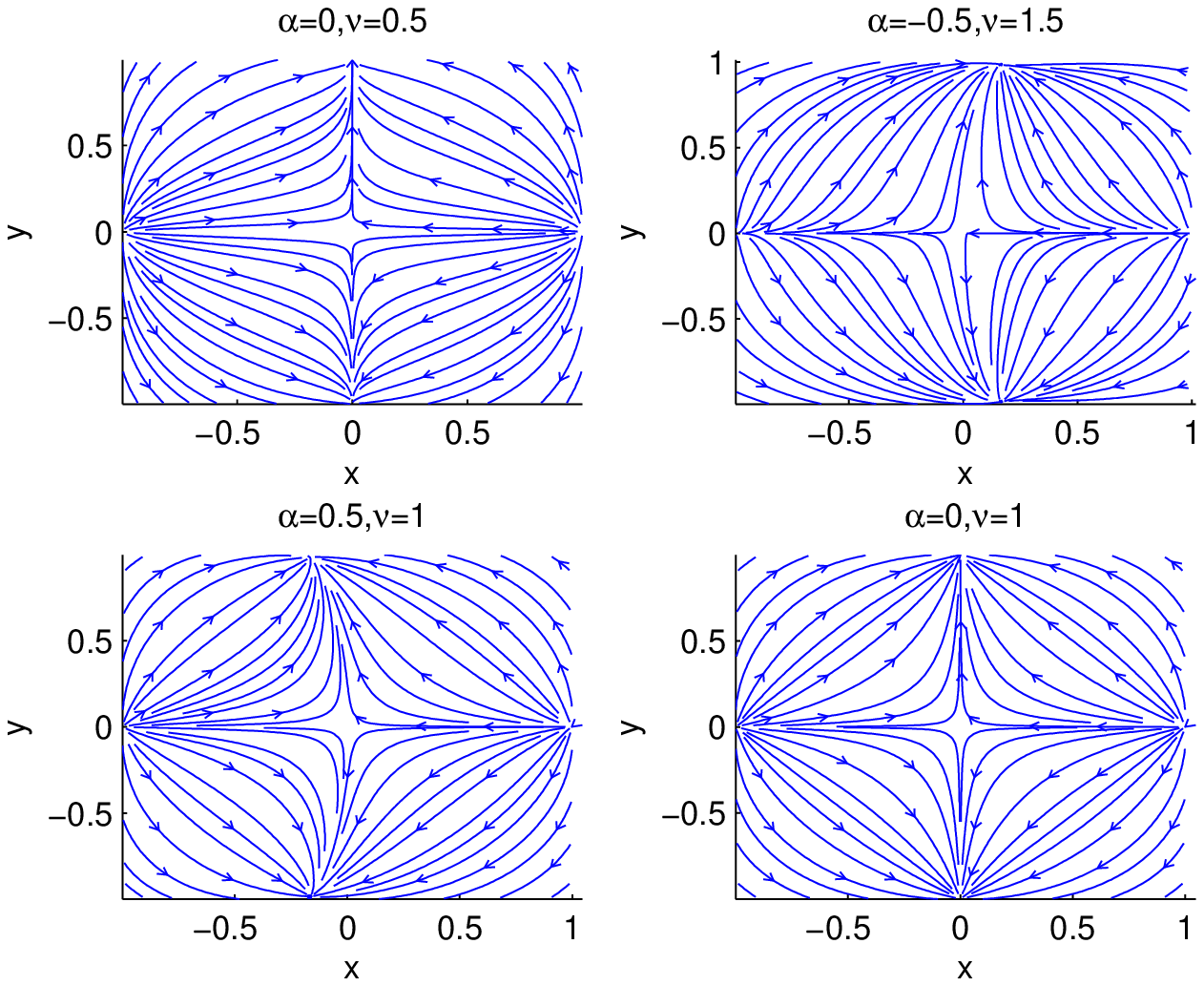}
\caption{$C_3$ }\label{fig23}
\end{center}
\end{figure}

\begin{figure}[H]
\begin{center}
\includegraphics
[scale=0.75]{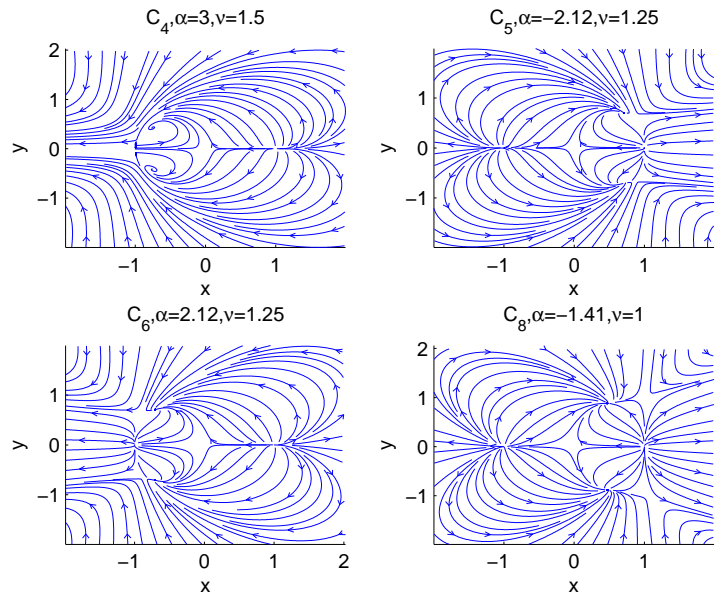}
\caption{$C_4,C_5,C_6,C_8$ }\label{fig24}
\end{center}
\end{figure}

\begin{figure}[H]
\begin{center}
\includegraphics
[scale=0.5]{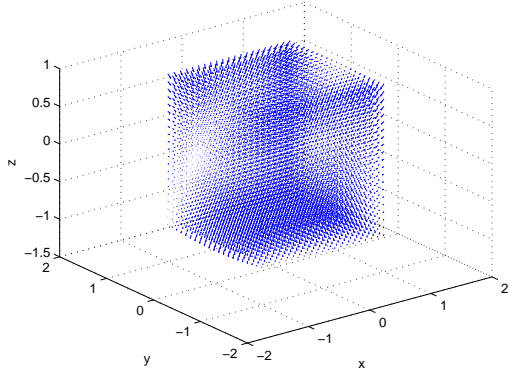}
\caption{Vector field diagram of (\ref{mads})}\label{fig25}
\end{center}
\end{figure}

\noindent At the end of this section, we will present the phase
plane diagrams of the autonomous system (\ref{mads}) for various
values of the parameter $\nu$ and $\alpha$ in figures
(\ref{fig21})-(\ref{fig24}) The last diagram (\ref{fig25}) depicts
the vector fields of the autonomous system (\ref{mads}).

%%%%%%%%%%%%%%%%%%%%%%%%%%%%%%%%%%%%%%%%%%%%%%%%%%%%%%%%%%%%%%%%%%%%%%%%%%%%%%%%%%%%%%%%%%%%%%%%%%%%%%%%%%%%%%%%%%%%%%%
\section{Cosmological Interpretations and Conclusion}\label{secIV}
%%%%%%%%%%%%%%%%%%%%%%%%%%%%%%%%%%%%%%%%%%%%%%%%%%%%%%%%%%%%%%%%%%%%%%%%%%%%%%%%%%%%%%%%%%%%%%%%%%%%%%%%%%%%%%%%%%%%%%%
\bigskip

\noindent The present work deals with a cosmological model
consisting of three non-interacting fluids namely the baryonic
matter in the form of perfect fluid ($p_b=(\nu-1)\rho_b$), dark
matter in the form of dust and dark energy as a scalar field
respectively. This cosmological model has been studied in the
framework of dynamical system analysis by forming the evolution
equations (Einstein's field equations) into an autonomous system
with suitable transformation of the variables. In table
\ref{tab1}, it has been shown that there are $10$ equilibrium
points ($C_1$-$C_{10}$) of the autonomous system. The values of
the relevant cosmological parameters at the equilibrium points
have also been presented in table \ref{tablast}.

\noindent The equilibrium point $C_1$ is completely dominated by
the baryonic matter and as expected in standard cosmology, the
model will be in decelerating phase if the baryonic fluid is
normal (ie, non-exotic: $\nu>\frac{2}{3}$) and it will be in the
accelerating era for exotic nature of the matter (ie,
$\nu<\frac{2}{3}$).

\noindent The equilibrium points $C_2$ and $C_4$ have identical
nature and both of them are not interesting from the cosmological
viewpoint as there is deceleration (of unusual magnitude) when
matter is completely dominated by DE. The point $C_3$ describes a
known result of standard cosmology: In the dust era there is
deceleration with the value of the deceleration parameter
$\frac{1}{2}.$

The equilibrium point $C_5$ and $C_6$ correspond to cosmological
era dominated by the dark energy(DE). The condition ($\alpha^2<3$)
restricts the DE to be exotic and there is acceleration. By
choosing $\alpha$ appropriately it is possible to match the model
with the recent observations.

\noindent Both the equilibrium points $C_7$ and $C_8$ in the phase
space have identical cosmological behaviors. For real points in
the phase space as well as for realistic baryonic fluid, $\nu$
should be restricted as $0<\nu<2.$ The cosmological model is
dominated by both the DE and the baryonic matter. In fact, the
cosmic phase described by these phase space points (ie $C_7$ and
$C_8$) has dominance of baryon over DE if $\frac{2}{3}<\nu<2$ and
there is deceleration. While if $0<\nu<\frac{2}{3}$ then DE
dominates the cosmic evolution and there is acceleration.

\noindent The scaling solutions represented by the equilibrium
points $C_9$ and $C_{10}$ are dominated by both the dark matter
(DM) and DE. But they are not of much physical interest as they
always correspond to dust era of evolution.

Thus, from the dynamical system analysis, the equilibrium points
describe various cosmological eras and some of them are
interesting from the cosmological viewpoints with recent observed
data. Therefore, one may conclude that from complicated
cosmological models one may get cosmological inferences without
solving the evolution equations, rather by analyzing with
dynamical system approach and the present work is an example of
this inference.

%%%%%%%%%%%%%%%%%%%%%%%%%%%%%%%%%%%%%%%%%%%%%%%%%%%%%%%%%%%%%

%%%%%%%%%%%%%%%%%%%%%%%%%%%%%%%%%%%%%%%%%%%%%%%%%%%%%%%%%%%%%


\section{References} \label{refer}
\begin{thebibliography}{26}


\bibitem{Perl99} S. J. Perlmutter {\it et al.} [Supernova Cosmology Project Collaboration], ``Measurements of Omega and Lambda from 42 high redshift supernovae'', {\it Astrophys. J.} {\bf 517} ({\bf 2}), 565-586(1999).
%doi:10.1086/307231
%[arXiv: astro-ph/9812133].\\

\bibitem{Rei98} A. J. Reiss {\it et al.} [Supernova Search Team],
``Observational evidence from supernovae for an accelerating universe and a cosmological constant'', {\it Astron. J.} {\bf 116} ({\bf 3}), 1009-1038(1998).
%doi:10.1086/300499
%[arXiv: astro-ph/9805201].\\

\bibitem{Per01} W. J. Percival {\it et al.} [2dFGRS Collaboration], ``The 2dF Galaxy Redshift Survey: The Power spectrum and the matter content of the Universe'', {\it Mon. Not. Roy. Astron. Soc.} {\bf 327} ({\bf 4}), 1297-1306(2001).
%doi:10.146/j.1365-8711.2001.04827.x
%[arXiv: astro-ph/0105252].\\

\bibitem{Sper07}
D. N. Spergel {\it et al.} [WMAP Collaboration], "Three Year Wilkinson Microwave Anisotropy Probe (WMAP) Observations : Implications for Cosmology", {\it Astrophys.\
J.\ Suppl.\ Ser.} {\bf 170} ({\bf 2}), 377-408 (2007).
%doi: 10.1086/513700
%[arXiv: astro-ph/0302209].\\

\bibitem{Teg04} M. Tegmark {\it et al.},"The Three Dimensional Power Spectrum of Galaxies from the Sloan Digital Sky Survey", {\it Astrophys.\ J.} {\bf 606} ({\bf 2}), 702-740(2004).
%doi:10.1086/382125
%[arXiv: astro-ph/0310723].\\

\bibitem{Eisen03} D. J. Eisenstein {\it et al.},{\it Astrophys. J.} {\bf 148}, 175(2003).
%doi:10.1086/466512
%[arXiv: astro-ph/0501171].\\

\bibitem{Komat09} E. Komatsu {\it et al.}[WMAP Collaboration], "FIVE-YEAR WILKINSON MICROWAVE ANISOTROPY PROBE OBSERVATIONS: COSMOLOGICAL INTERPRETATION", {\it Astrophys. J. Suppl. Ser.} {\bf 180} ({\bf 2}),
330-376(2009).
%doi:10.1088/0067-0049/180/2/330
%[arXiv: 1001.4538 [astro-ph.CO]].\\

\bibitem{Wein89} S. Weinberg,"The Cosmological Constant Problem", {\it Rev. Mod. Phys.} {\bf 61} ({\bf 1}), 1-23(1989).
%doi:10.1103/RevModPhys.61.1\\

\bibitem{AT10} L. Amendola, S. Tsujikawa, "Dark Energy" : Theory.
and Observations'', Cambridge, UK\,: Cambridge University Press (2010).%\\

\bibitem{MC15}
N.~Mahata and S.~Chakraborty, "A Dynamical system analysis of
three fluid cosmological model",
%doi:
[arXiv:1512.07017v1[gr-qc]], 22 Dec,2015.%\\

\bibitem{Carrol01} S. M. Carroll,{\it Liv. Rev. Lett.} {\bf 4}, 1(2001).%\\

\bibitem{Wang10} Y. Wang, "Dark Energy", Willey-VCH (2010).%\\

\bibitem{Per91} L. Perko, "Differential Equations and Dynamical Systems", Springer-Verlag : New York (1991).%\\

\bibitem{AP90} D. K. Arrowsmith and C. M. Place, "An Introduction to Dynamical Systems", Cambridge Univ. Press : Cambridge, England (1990).%\\

\bibitem{Wig03} S. Wiggins, "Introduction to Applied Nonlinear Dynamical Systems and Chaos", "2nd Edition", Springer, Berlin (2003).%\\

\bibitem{PR03} P. J. E. Peebles and B. Ratra, "The cosmological constant and dark energy", {\it Rev. Mod. Phys.} {\bf 75}({\bf 2}),
559-606(2003).
%doi: 10.1103/revmodphys.75.559\\

\bibitem{RP88} B. Ratra and P. J. E. Peebles, "Cosmological consequences of a rolling homogeneous scalar field", {\it Phys. Rev. D.} {\bf 37} ({\bf 12}), 3406-3427(1988).
%doi: 10.1103/physrevd.37.3406\\

\bibitem{CDS98} R. R. Caldwell and R. Dave and P. J. Steinherdt, "Cosmological Imprint of an Energy Component with General Equation of State", {\it Phys. Rev. Lett.} {\bf 80}({\bf 8}),
1582-1585(1998).
%doi: 10.1103/PhysRevLett.80.1582\\

\bibitem{AMS00} C. Armendariz-Picon and V. Mukhanov and P. J. Steinherdt, "Dynamical Solution to the Problem of a Small Cosmological Constant and Late-Time Cosmic Acceleration", {\it Phys. Rev. Lett.} {\bf 85}({\bf 21}),
4438-4441(2000).
%doi: 10.1103/PhysRevLett.85.4438%\\

\bibitem{ASB} Amartya S. Banerjee, "An Introduction to Center Manifold Theory".%\\

\bibitem{AMS01} C. Armendariz-Picon and V. Mukhanov and P. J. Steinherdt, "Essentials of k-essence", {\it Phys. Rev. D.} {\bf 63}({\bf 10}),
103510(2001).
%doi: 10.1103/PhysRevD.63.103510\\

\bibitem{CKW03} R. R. Caldwell and M. Kamionkowski and N. N. Weinberg, "Causes a Cosmic Doomsday", {\it Phys. Rev. Lett.} {\bf 91}({\bf 7}),
071301(2003).
%doi: 10.1103/physrevlett.91.071301\\

\bibitem{S02} A. Sen,{\it  J. H. E. P} {\bf 207},
65(2002).%\\

\bibitem{CST06} E. J. Copeland and M. Sami and S. Tsujikawa, "DYNAMICS OF DARK ENERGY", {\it Int. J. Mod. Phys. D} {\bf 15}({\bf 11}),
1753-1935(2006).
%doi: 10.1142/S021827180600942X\\

\bibitem{TSAM12} M. Tsamparlis and A. Paliathanasis,"Three-fluid cosmological model using Lie and Noether symmetries", {\it Class. Quantum Gravity.} {\bf 29}({\bf 1}),
015006(2012).
%doi: 10.1088/0264-9381/29/1/015006\\

\bibitem{CLW98} E. J. Copeland and A. R. Riddle and D. Wands, "Exponential potentials and cosmological scaling solutions.", {\it Phys. Rev. D.} {\bf 57}({\bf 8}),
4686-4690(1998).
%doi: 10.1103/PhysRevD.57.4686\\


\end{thebibliography}
\end{document}